\documentclass[12pt,a4paper]{article}
\usepackage[latin1]{inputenc}
\usepackage{amsmath}
\usepackage{amsfonts}
\usepackage{amssymb}
\usepackage{graphicx}
\usepackage[left=2cm,right=2cm,top=2cm,bottom=2cm]{geometry}
\usepackage{setspace}
\usepackage[square,sort,comma,numbers]{natbib}

\newcommand{\alp}{$\alpha$}
\newcommand{\bet}{$\beta$}

\begin{document}
\bibliographystyle{unsrt}

\title{Learning from history: Non-Markovian analyses of complex trajectories for extracting long-time behavior}
\author{Ernesto Suarez and Daniel Zuckerman\\
\\
Department of Computational and Systems Biology, University of Pittsburgh
\\}
\date{\today}
\maketitle

\onehalfspacing
\section*{Abstract}

A number of modern sampling methods probe long time behavior in complex biomolecules using a set of relatively short trajectory segments.  Markov state models (MSMs) can be useful in analyzing such data sets, but in particularly complex landscapes, the available trajectory data may prove insufficient for constructing valid Markov models.  Here, we explore the potential utility of history-dependent analyses applied to relatively poor decompositions of configuration space for which MSMs are inadequate.  Our approaches build on previous work [Suarez et. al., JCTC 2014] showing that, with sufficient history information, unbiased equilibrium and non-equilibrium observables can be obtained even for arbitrary non-Markovian divisions of phase space.  We explore a range of non-Markovian approximations using varying amounts of history information to model the finite length of trajectory segments, applying the analyses to toy models as well as several proteins previously studied by $\mu$sec$-$msec scale atomistic simulations [Lindorff-Larsen et. al., Science 2011].

\section{Introduction}

The extremely complex dynamics of biomolecules can be difficult to sample and understand using straightforward molecular dynamics simulations, motivating the popularity of Markov state models (MSMs) which have been successfully applied to a number of systems \cite{Bowman2010,Chodera2007,Hinrichs2007,Beauchamp2011,Beauchamp2012}.  However, MSMs can require extensive sampling and the careful definition of many states, and even so, the resulting modeling may exhibit non-Markovian behavior \cite{Beauchamp2012}.

The question we investigate here is whether non-Markov models and analyses may be useful in some cases.  We have shown in recent work \cite{Suarez2014} that using a subset of history information in the analysis of highly non-Markovian states is sufficient to build a model that gives unbiased state probabilities and kinetic properties. It is enough to introduce in the rate-matrix formulation and for each macroscopic transition, the last state visited. In other words, with sufficient history information, it is not necessary to seek an optimal partitioning of the space or optimal slow coordinates as order parameters.

The main goal of this work is explore the potential utility of non-Markovian analyses of trajectories.  We try to obtain good kinetic estimates not only for arbitrary non-Markovian decompositions of configuration space, but also when the history information is limited in the sense that it is not possible to unambiguously assign the last state visited by a trajectory.  That is, although we analyze very long trajectories, we consider non-Markovian approximations using varying amounts of history information to model the situation of finite trajectory segments. The analyses are applied to toy models as well as to several proteins previously studied by $\mu$sec$-$msec scale atomistic simulations \cite{Lindorff-Larsen2011}.

\section{Theoretical formulation}\label{sec:methods}

We compare a number of different analyses to well-sampled ``brute force'' trajectories, which provide reliable reference results. The focus here is on ``poor'' decompositions of configuration space which lead to non-Markovian behavior by construction.

\subsection*{Reference calculation of observables from long trajectories}

All the simulations presented here are single regular brute force (BF) trajectories. The populations of configuration-space regions or ``bins'' are obtained by the trivial estimator $\hat{p}_i=c_i/C$, that turns out to be the maximum likelihood estimator (MLE), where $c_i$ is the number of times the systems was in bin $i$, and $C$ is the total number of configurations. The first passage times (MFPTs) are measured and averaged directly during the simulation by just following the evolution of the trajectories.

\subsection*{Markovian calculation of observables}
A traditional Markov analysis of the trajectories is performed for reference; this would be the case where no history information is included in the matrix analysis. The key quantity of interest is the presumed history-independent rate $k_{ij}$ between two bins, defined by the conditional probability 
\begin{equation}\label{eq:ratedef}
k_{ij}=P\{X_{t+\tau}=j|X_t=i\},
\end{equation}
where $X_t$ is the random variable representing the state of the system at time $t$, and $\tau$ is the lag-time used for
 the Markov model.  Each rate is estimated by the MLE estimator

\begin{equation}\label{rateestimation}
\hat{k}_{ij}=c_{ij}/c_i.
\end{equation}

Here we are not using any constraints in the estimation, such as the one used for symmetric matrices \cite{Beauchamp2011}, since the rates are reasonably well sampled in our analyses as can be inferred by comparing the matrix results with direct BF measurements. The MFPTs are computed analytically as shown in reference \cite{Suarez2014}.

\subsection*{Fully-history non-Markovian calculation of observables}
This method has been previously described in detail \cite{Suarez2014} and uses the full history to label the trajectories depending of which is the last state visited.
Suppose we have only two macroscopic states of interest A and B, which generally may not cover the full phase space.  Every segment of a trajectory can be given a label according to whether the system was last in state A (the label $\alpha$) or B (label $\beta$).  

The labeled matrix approach explicitly uses the decomposition of the equilibrium population into $\alpha$ and $\beta$ component for each bin $i$:

\begin{equation}\label{probcolor}
p^{eq}_i=p^{\alpha}_i+p^{\beta}_i.
\end{equation}

Then, with $N$ bins, a set of $2N$ probabilities is required. Similarly, a $2N\times2N$ rate matrix is used for the analysis, and Eq.~\ref{rateestimation} is transformed into

\begin{equation}\label{ratecolorestimation}
\hat{k}^{\mu}_{ij}=c^{\mu}_{ij}/c^{\mu}_i, \;\;\; \mu=\alpha,\beta,
\end{equation}
where $\mu$ can be either label $\alpha$ or $\beta$ depending on the last state visited. The bin probabilities and the MFPTs can be solved analytically \cite{Suarez2014}.

\subsection*{Limited history analysis: Second-order Markov approach}
We are also interested in modeling the case where a large set of unbiased trajectory segments is generated, perhaps from distributed computing or from a replica exchange simulation.  To this end, we consider non-Markov analyses that employ time-limited history.  Thus, for this particular analysis, we assume it is \emph{not} possible to label all the trajectories -- in the case there is more than one -- as \alp\ or \bet .
 
Under limitation of finite history, a first approach could be to increase the order of the Markov approximation. The rates in a second order Markov model have three indexes, since they will depend on the state -- in this case, the bin -- the system occupied at times $t$ and $t-\tau$ :

\begin{equation}\label{2ndOrderMarkov}
k_{ij|m}=P\{X_{t+\tau}=j|X_t=i,X_{t-\tau}=m\}.
\end{equation}
Similarly, the rates $\hat{k}_{ij|m}=c_{ij|m}/c_{i|m}$ are estimates from the transition counts, taking into account $m$, the bin occupied at $t-\tau$.

Numerical estimates of observables are obtained by kinetic simulation of the rate matrix.

\subsection*{Limited history analysis: Partial-color approach}
We know that if it is possible to assign a ``color'' (label as $\alpha$ or $\beta$) to all the trajectories, the non-Markovian formulation is an unbiased way to estimate observables \cite{Suarez2014}. The main point of the partial-color method is to take advantage of the labeled information if it is available.  Below, we examine analyses based on an amount of history equal to 5\% and 10\% of the average MFPT (average over forward and reverse directions).  That is, when examining a given time point of the trajectory for estimating a labeled rate, the $\alpha$ or $\beta$ label is only assigned properly if the trajectory occupied either of states A or B (as opposed to the intermediate region) during the preceding 5 or 10\% of the trajectory.

Construction of the rate matrix proceeds in two steps.  The first step is to build and solve a $2N \times 2N$ rate matrix where the rates are color independent, $k^\mu_{ij}=k_{ij}$, since that provides the Markovian values of $p^\alpha_i$ and $p^\beta_i$ as well as default rate values.  Next, each point in the trajectory is examined.  If there is sufficient history to assign an $\alpha$ or $\beta$ label, that is done.  If not, the default Markovian values are used, in effect: the label $\mu$ is assigned to probability $p^\mu_i$.  This procedure is used to build the non-Markovian (labeled) $2N \times 2N$ count matrix $\boldsymbol{C}=\{c^\mu_{ij}\}$, that is transformed into a rate matrix (Eq.~\ref{ratecolorestimation}) used for the estimation of the observables.

\section{Model systems and simulation details}

We studied long trajectories for and protein folding systems generated by Shaw and coworkers \cite{Lindorff-Larsen2011}, as well as simple toy models.

\subsection{Toy models}
\begin{figure}[ht]
    \centering
    \includegraphics[width=0.5\textwidth]{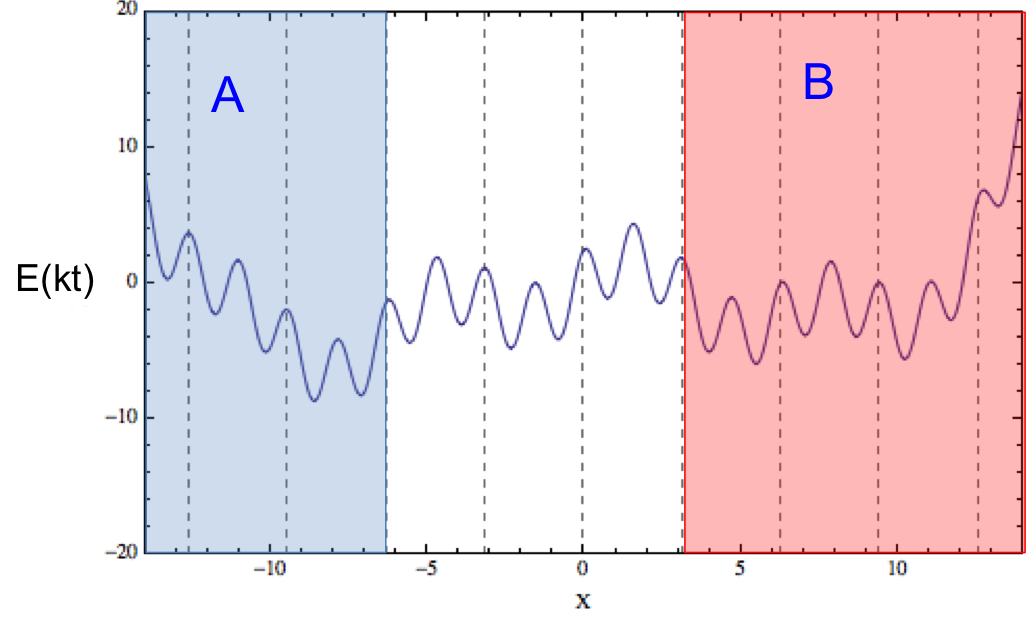}
    \caption{1D toy model. The figure shows the definition of the states A and B and the partition of the space in bins is indicated with dashed lines.}
    \label{fig:1Dtoy}
\end{figure}

Monte Carlo (MC) simulations where performed on two different toy models. The first is a one-dimensional model represented in Fig.~\ref{fig:1Dtoy}. The space is divided in 10 bins, most of them of width $\pi$. The energy funtion in $k_BT$ units is given by 

\begin{equation}
E_{1D}=\left\{\begin{array}{cl}
\sin(x)+2.5\cos(4x)+0.0008x^4-0.11(x - 0.5)^2 &\mbox{ if $-14<x<14$} \\
\infty &\mbox{ otherwise}
\end{array} \right.
\end{equation}

Two states A and B are defined as shown in Fig.~\ref{fig:1Dtoy}.  The state A consists of the first three bins while state B is the last four bins.

About $10^6$ MC iterations were done for this model where the trial move $\delta x$ is chosen randomly in the interval $[-\pi/2,\pi/2]$ with uniform probability distribution. Notice that the average displacement in $x$ is $\pi/4$, four times smaller than the bin size.

\begin{figure}[ht]
    \centering
    \includegraphics[width=0.8\textwidth]{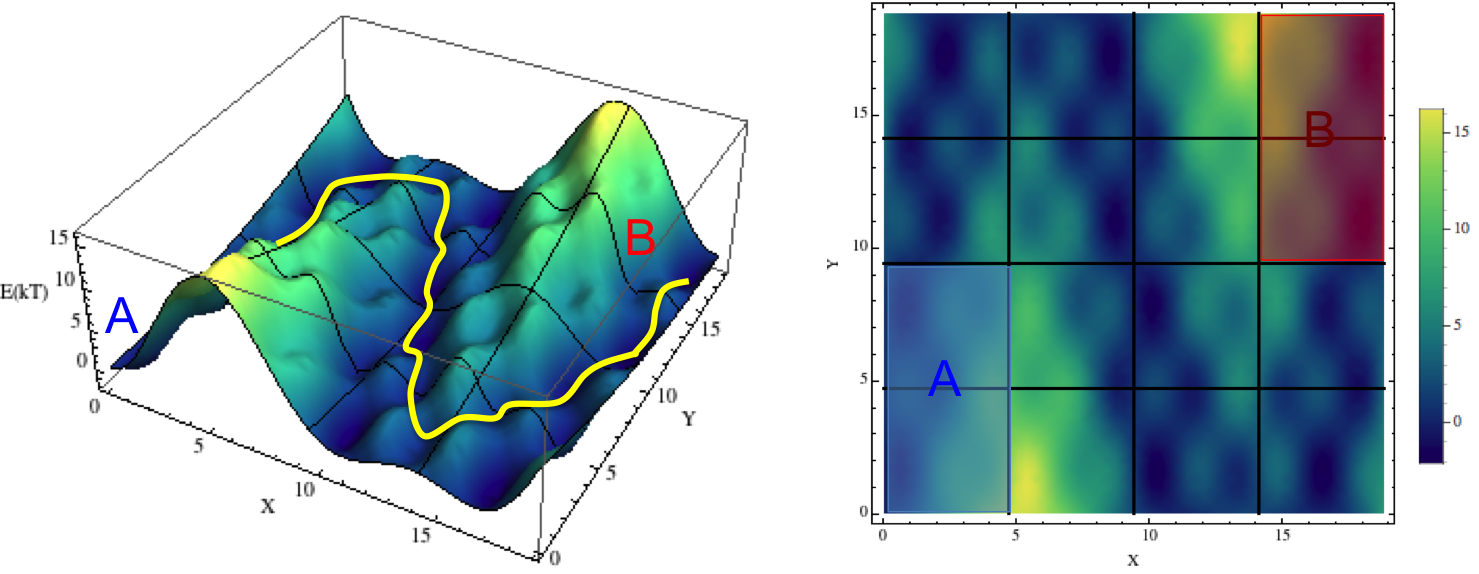}
    \caption{2D toy model. Representation of the energy function and a possible path between A and B (left). Density plot indicating the partitioning of the space in bins and the definition of the states A and B in terms of bins (right). }
    \label{fig:2Dtoy}
\end{figure}

The second model is the two-dimensional system shown in Fig.~\ref{fig:2Dtoy}.  The states A and B are defined by two bins in the configurational space, and the total number of bins used was 16 (see Fig.~\ref{fig:2Dtoy} right), each bin with dimensions $1.5\pi \times 1.5\pi$. The energy function in $k_BT$ units is given this time by Eq.~\ref{eq:2Dtoy}. 

\begin{equation}\label{eq:2Dtoy}
E_{2D}=\left\{\begin{array}{cl}
-4.2\sin(2x)\sin(y)+0.02(y - (23\sin(x/3) + x))^2 &\mbox{ if $0<x,y<6\pi$} \\
\infty &\mbox{ otherwise}
\end{array} \right.
\end{equation}

The number of MC steps performed was about $2\times10^6$, and the components of the trial moves $\delta x$ and $\delta y$ where selected independently and with uniform distribution in the interval $[-3\pi/8,3\pi/8]$.

A second partitioning of the space was also considered in order to create what we call pseudo one-dimensional model. In Fig.~\ref{fig:2Dtoy} there is a representation of what would be an ``optimal path'', and the new binning tries to make bins almost perpendicular to that path, but of course, not in an optimal way. The goal here is to create a harder system to test the success of the methods described in Section \ref{sec:methods}.

\begin{figure}[ht]
    \centering
    \includegraphics[width=0.45\textwidth]{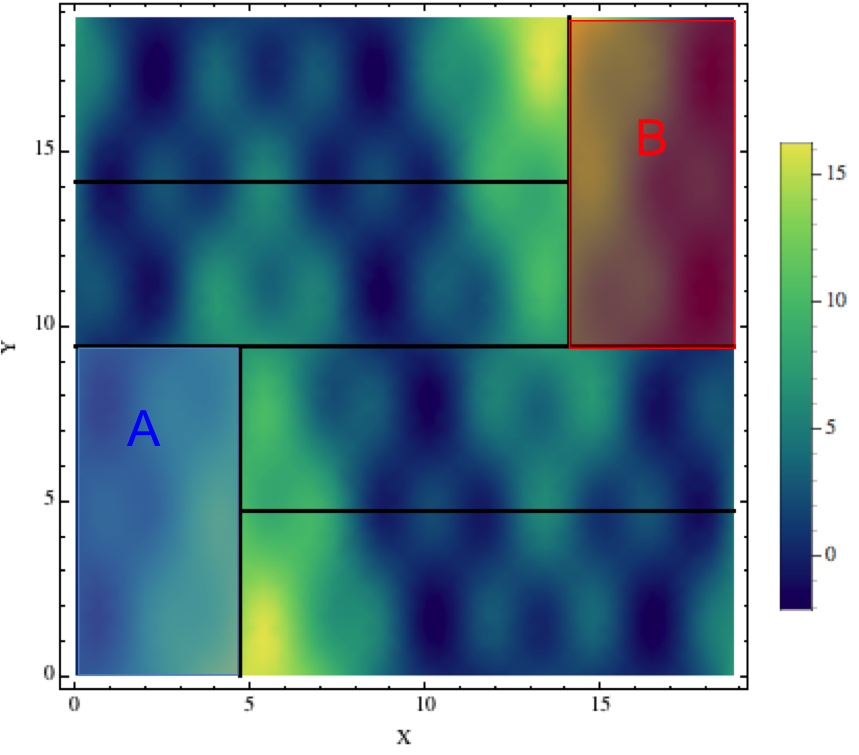}
    \caption{Pseudo 1D toy model}
    \label{fig:pseudo1Dtoy}
\end{figure}\

\subsection{Protein Models}
Trajectories for five protein previously studied in explicit solvent by $\mu$sec$-$msec scale atomistic molecular dynamics simulations \cite{Lindorff-Larsen2011}, were also analyzed using the approaches described in Section \ref{sec:methods}.  We examined Chignolin, Trp cage, BBA, NTL9, and Villin.  In the cases of BBA and NTL9, more than one trajectory was available, but we only analyzed the longest trajectory reported for each protein.

To create non-Markovian bins, the alpha-carbon RMSD (compared to the folded structure, excluding terminal residues) was used to define bins for the matrix analyses for all the proteins.  All the trajectories were saved every $\tau$=0.2ns, and we used this same $\tau$ for the matrix analyses. 

Table \ref{table:proteins} shows, for each system, the definition of the states A (folded) and B (unfolded), as well as the number of bins used, the total simulation time considered for the analysis and the number of residues.  Further details about the trajectories can be found in the original paper \cite{Lindorff-Larsen2011}.

\begin{table}[ht]
\caption{Protein models used for Markovian and non-Markovian analyses. For each system, the table shows the number of residues, the total simulation time considered, the number of bins for the analyses and the states definitions.
\label{tab:binsStates}
} 
\centering 
\begin{tabular}{l c c c c c} 
\hline\hline 
Protein & Num.\,Residues & Time($\mu s$) & Num.\,Bins & RMSD\,State\,A  & RMSD\,State\,B \\ [1.0ex] 
\hline 
Chignolin & 10 & 106  & 16 &  $<2.0$\AA & $>5.0$\AA \\ 
Trp-cage  & 20 & 208  & 9  &  $<1.4$\AA & $>6.0$\AA \\ 
BBA       & 28 & 225  & 6  &  $<3.0$\AA & $>7.0$\AA \\ 
NTL9      & 39 & 1100 & 7  &  $<2.0$\AA & $>5.0$\AA \\ 
Villin    & 35 & 125  & 10 &  $<2.0$\AA & $>6.0$\AA \\ [1ex] 
\hline 

\end{tabular} 
\label{table:proteins} 
\end{table}

\section{Results}
In this section we will show compare performance of different types of matrix analysis in terms of the MFPTs. The bin populations are also estimated from the matrices, but are not challenging to estimate from a Markovian formulation.  The stationary distribution $\mathbf{p}$ from a Markov model obeys $K^T\mathbf{p}=\mathbf{p}$ and is equivalent to imposing the steady-state condition
\begin{equation}
\frac{d\mathbf{p}}{dt}=(K^T-I)\mathbf{p}=0,
\end{equation}
where $K$ is the rate matrix and $I$, the identity matrix.

\subsection{Toy models}
The MFPTs were measured directly from the brute force trajectories, and also estimated from the matrix analyses described in Section \ref{sec:methods}.  Since we are only interested in how the matrix analyses perform with respect to the direct BF measurement we divide all MFPT values by the corresponding BF direct measurement.

\begin{figure}[ht]
    \centering
    \includegraphics[width=0.6\textwidth]{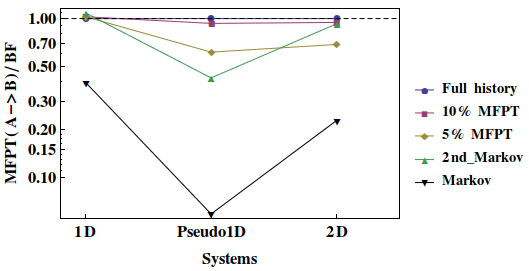}
    \caption{Relative MFPT from A to B, obtained from matrix analyses for the 1D, pseudo 1D, and the 2D models. The estimates are done using the full-history (non-Markovian), partial color analysis with 5\% and 10\% of (MFPT(AB)+MFPT(BA))$/2$, and second order Markov approximation.}
    \label{fig:MFPTABtoys}
\end{figure}

\begin{figure}[ht]
    \centering
    \includegraphics[width=0.6\textwidth]{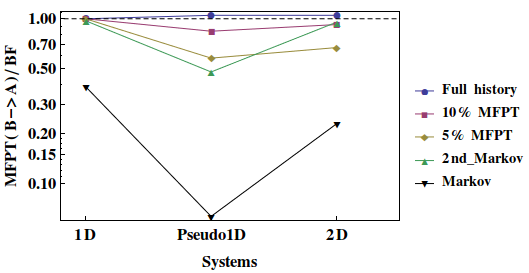}
    \caption{Relative MFPT from B to A, obtained from matrix analyses for the 1D, pseudo 1D, and the 2D models. The estimates are done using the full-history (non-Markovian), partial color analysis with 5\% and 10\% of (MFPT(AB)+MFPT(BA))$/2$, and second order Markov approximation.}
    \label{fig:MFPTBAtoys}
\end{figure}

Fig.~\ref{fig:MFPTABtoys} and \ref{fig:MFPTBAtoys} show, respectively, the relative MFPTs from A to B and from B to A for the three toy systems: the one-dimensional toy model, the so-called pseudo one-dimensional model and the two-dimensional one. 

In the plots legends ``full history'' means we are doing non-Markovian analysis, 
where we have enough history to label all of the trajectory according to the last of states A or B visited. On the other hand, ``$x$\% MFPT'' indicates partial color analysis where we are using only a history size of $x$\% of (MFPT(AB)+MFPT(BA))$/2$. Finally, ``2nd\_Markov'' is a second-order Markov approximation.

\subsection{Protein models}
\begin{figure}[ht]
    \centering
    \includegraphics[width=0.6\textwidth]{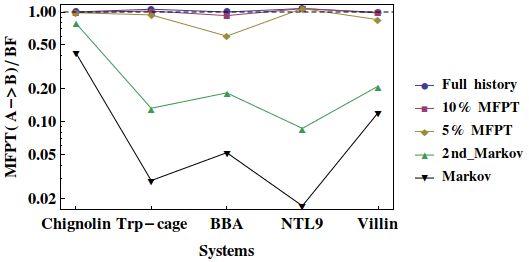}
    \caption{Relative MFPT from A to B, obtained from matrix analyses for protein models. The estimates are done using the full-history (non-Markovian), partial color analysis with 5\% and 10\% of (MFPT(AB)+MFPT(BA))$/2$, and second order Markov approximation}
    \label{fig:MFPTABprots}
\end{figure}

\begin{figure}[ht]
    \centering
    \includegraphics[width=0.6\textwidth]{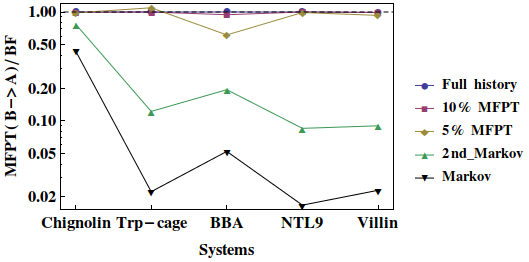}
    \caption{MFPTAB Relative MFPT from B to A, obtained from matrix analyses for protein models. The estimates are done using the full-history (non-Markovian), partial color analysis with 5\% and 10\% of (MFPT(AB)+MFPT(BA))$/2$, and second order Markov approximation}
    \label{fig:MFPTBAprots}
\end{figure}

We applied the same analyses to the long-timescale trajectories generated by Shaw and coworkers \cite{Lindorff-Larsen2011}.  As in the previous section, all the MFPT values plotted here are relative to the direct BF measurements. Fig.~\ref{fig:MFPTABprots} and \ref{fig:MFPTBAprots} show, respectively, the relative MFPTs from A to B and from B to A, obtain from  for the five protein models considered. 

Again, ``full history'' means we are doing non-Markovian analysis, where we have enough history to label all time-points in trajectory, ``$x$\% MFPT'' is a partial color analysis where only a history size of $x$\% of (MFPT(AB)+MFPT(BA))$/2$ is taken into account, and ``2nd\_Markov'' is a second order Markov approximation.

\section{Discussion and Conclusions}
How can trajectory data be analyzed in the absence of a highly Markovian decomposition of phase space?  This report presents an initial exploration of non-Markovian analyses applied to a wide range of systems, from toy models to proteins, in cases where phase-space decompositions were highly non-Markovian by construction.  Perhaps the main conclusion is that a little history goes a long way: even a second-order Markov model with a short lag time provides reasonable estimates of first-passage times.  An estimate can probably be considered reasonable if it is roughly within a factor of $e$ from the true value, suggesting order $k_B T$ error in the effective barrier height --- which is well within the accuracy limit of modern force fields \cite{Shirts2005a,Pande-2003c}.  If further history information can be included, such as whether a trajectory previously visited one of the A or B macrostates of interest for rate estimation, better estimates can be obtained.

Although these results are encouraging, further studies should help to elucidate the general utility of these analyses.  For example, how well do the analyses perform on less complete trajectories?  At the same time, superior non-Markovian analyses may be within reach.  It should not be difficult, for instance, to construct higher-order Markov models by employing progressively coarser decompositions of phase space further back into the history of a trajectory.  Also, additional labeling information, including intermediate states/bins, may prove useful.

\section*{Acknowledgements}
We thank Joshua Adelman for helpful discussions and the NSF for support (Grant Nos. MCB-0643456, MCB-1119091 and MCB- 0845216).

\bibliography{dmz}
\end{document}